\documentclass[iop]{emulateapj}
\usepackage{apjfonts}

\begin{document}

\title{The Blue Hook Populations of Massive Globular Clusters\altaffilmark{1}}

\author{Thomas M. Brown\altaffilmark{2}, 
Allen V. Sweigart\altaffilmark{3},
Thierry Lanz\altaffilmark{4},
Ed Smith\altaffilmark{2}, 
Wayne B. Landsman\altaffilmark{5},
Ivan Hubeny\altaffilmark{6}
}

\altaffiltext{1}{Based on observations made with the NASA/ESA Hubble
Space Telescope, obtained at STScI, and associated with proposal
10815.}

\altaffiltext{2}{Space Telescope Science Institute, 3700 San Martin Drive,
Baltimore, MD 21218;  tbrown@stsci.edu, edsmith@stsci.edu} 

\altaffiltext{3}{Code 667, NASA Goddard Space Flight Center, Greenbelt, MD
20771; allen.v.sweigart@nasa.gov}

\altaffiltext{4}{Department of Astronomy, University of Maryland, College
Park, MD 20742; lanz@astro.umd.edu}

\altaffiltext{5}{Adnet Systems, NASA Goddard Space
  Flight Center, Greenbelt, MD 20771; wayne.b.landsman@nasa.gov}

\altaffiltext{6}{Steward Observatory, University of Arizona, Tucson,
  AZ 85712; hubeny@aegis.as.arizona.edu}

\submitted{To appear in The Astrophysical Journal}

\begin{abstract}

We present new {\it HST} ultraviolet color-magnitude diagrams of 5
massive Galactic globular clusters: NGC~2419, NGC~6273, NGC~6715,
NGC~6388, and NGC~6441. These observations were obtained to
investigate the ``blue hook'' phenomenon previously observed in UV
images of the globular clusters $\omega$ Cen and NGC~2808. Blue hook
stars are a class of hot (approximately 35,000~K) subluminous
horizontal branch stars that occupy a region of the HR diagram that is
unexplained by canonical stellar evolution theory. By coupling new
stellar evolution models to appropriate non-LTE synthetic spectra, we
investigate various theoretical explanations for these
stars. Specifically, we compare our photometry to canonical models at
standard cluster abundances, canonical models with enhanced helium
(consistent with cluster self-enrichment at early times), and flash-mixed
models formed via a late helium-core flash on the white dwarf cooling
curve. We find that flash-mixed models are required to explain the
faint luminosity of the blue hook stars, although neither the
canonical models nor the flash-mixed models can explain the range of
color observed in such stars, especially those in the most metal-rich
clusters.  Aside from the variation in the color range, no clear trends
emerge in the morphology of the blue hook population with respect
to metallicity.

\end{abstract}

\keywords{globular clusters: general -- globular clusters: individual
  (NGC~2419, NGC~6273, NGC~6715, NGC~2808, NGC~6388, NGC~6441) -- stars:
  atmospheres -- stars: evolution -- stars: horizontal branch --
  ultraviolet: stars}

\section{Introduction}

Understanding the formation history and chemical evolution of the
Galactic globular clusters has always been a formidable, but
fundamental, task.  However, this task has become even more
challenging with the recent discovery of complex stellar populations within
individual globular clusters.  The most well-known example is
$\omega$~Cen (Anderson 1997; Bedin et al.\ 2004; Ferraro et
al.\ 2004), but the same phenomenon has also been discovered in other
massive globular clusters, including NGC~2808 (D'Antona et al.\ 2005;
Piotto et al.\ 2007), NGC~1851 (Milone et al.\ 2008; Han et
al.\ 2009), NGC~6715 (M54; Layden \& Sarajedini 2000; Siegel et
al.\ 2007), and NGC~104 (47~Tuc; Anderson et al.\ 2009).  In
$\omega$~Cen, accurate photometry demonstrates that its main sequence (MS)
population is split into multiple sequences, with the bluest MS stars
being more metal rich, implying that the corresponding sub-population
is perhaps enhanced in helium ($Y \sim 0.4$; Piotto et al.\ 2005).
Such helium-rich stars may have formed in a second stellar generation from
the helium-rich ejecta of the initial stellar generation that formed the
cluster (e.g., see Renzini 2008).  Possible candidates for producing
these ejecta include asymptotic-giant-branch (AGB) stars, massive
MS stars, type II supernovae, and even stars external to
the cluster (Bekki \& Norris 2006 and references therein).  These
clusters thus refute the long-standing premise that globular
clusters are simple stellar populations.

The discovery of complex populations in globular clusters has
understandably generated much excitement, especially in
regard to the morphology of the horizontal branch
(HB).  Because the MS turnoff (MSTO) mass decreases strongly with
increasing helium at a fixed cluster age, a high helium abundance will
lead to a bluer HB morphology for a given range in the red-giant branch
(RGB) mass loss, and this, in turn, might account for the long
blue HB tails found
in some clusters (D'Antona et al.\ 2002; Busso et al.\ 2007).  In this
scenario, the extreme HB (EHB) stars, located at the hot end of the HB
at temperatures $T_{\rm eff} \gtrsim 16,000~K$ and surface gravities log~$g
> 5$, would be the progeny of the most helium-rich MS stars.  The hottest
EHB stars have extremely thin envelope masses ($\sim
10^{-3}~M_\odot$), due to extensive mass loss on the RGB.
The analogs of the EHB stars in the field are the subdwarf B (sdB)
stars.  Both unresolved spectroscopy and resolved imaging have
demonstrated that EHB stars are the dominant contributors to the ``UV
upturn'' in the otherwise cool spectra of elliptical galaxies (Brown
et al.\ 1997; Brown et al.\ 2000; Brown et al.\ 2008).

\begin{figure*}[t]
\epsscale{1.1}
\plotone{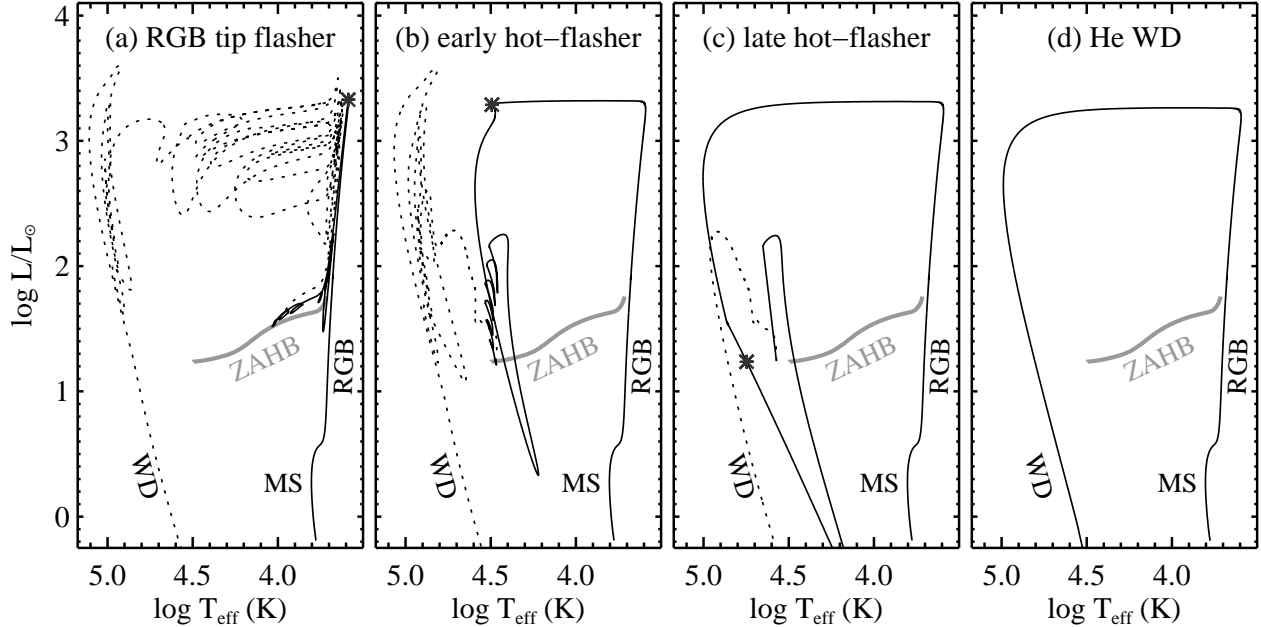}
\epsscale{1.1}
\caption{ Various evolutionary paths for producing an HB
  star. The zero-age HB (ZAHB) phase is highlighted in grey, while the
  pre-ZAHB evolution ({\it solid curves}) and post-ZAHB evolution
  ({\it dashed curves}) are in black. The peak of the helium-core flash is marked by
  an asterisk. The four panels show the evolution for progressively
  larger amounts of mass loss on the RGB. The evolution in the first
  two panels produces canonical HB stars in which the H-rich surface
  composition does not change during the helium-core flash. In the third
  panel, the helium-core flash occurs on the WD cooling curve, producing a
  flash-mixed star having a surface composition highly enriched in 
  helium and carbon and a temperature significantly hotter than the canonical HB. 
  In the fourth panel, the helium flash never occurs and the star dies as a helium WD.}
\end{figure*}

UV observations of globular clusters have revealed yet another
puzzling result, namely, that two of the massive clusters with
evidence for a helium-rich MS population ($\omega$ Cen and NGC~2808) also
contain an unexpected population of subluminous EHB stars that cannot
be explained by canonical stellar evolution theory.  These subluminous
stars were first discovered in $\omega$~Cen (D'Cruz et al.\ 1996,
2000), where they form a ``blue hook'' (BH) feature at the hot end of
the canonical EHB.  D'Cruz et al.\ (1996) proposed that such stars may
have undergone a delayed helium-core flash after peeling away from the RGB
due to high mass loss.  

We now know that stars can follow a variety
of evolutionary paths to the HB, depending on when the helium-core flash
occurs (see Figure 1).  For normal rates of mass loss, the helium-core
flash will occur at the tip of the RGB (RGB tip flasher, Figure 1a).
However, for sufficiently high rates of mass loss, the helium-core flash can occur
either during the crossing of the HR diagram (early hot flasher,
Figure 1b) or on the white dwarf (WD) cooling curve (late hot
flasher, Figure 1c), as first shown by
Castellani \& Castellani (1993).  Brown et al.\ (2001; hereafter
Paper I) demonstrated that a delayed helium-core flash on the WD
cooling curve would result in a stellar atmosphere
extremely enhanced in helium ($\sim$96\% by mass) and carbon
($\sim$4\% by mass), due to the mixing of the stellar envelope into the
hot helium-burning core.  This ``flash mixing'' decreases the
opacity below the Lyman limit, thereby lowering the flux at longer
wavelengths, and increases the effective temperature
in the EHB stars.  In Paper I, we found that both of these effects
together might explain the BH feature in the color-magnitude diagram
(CMD) of NGC~2808.  Evidence for BH populations in other
clusters soon followed, including NGC~6715 (Rosenberg et al.\ 2004),
NGC~6388 (Busso et al.\ 2007), NGC~2419 (Ripepi et al.\ 2007), and
NGC~6273 (noted by Rosenberg et al.\ 2004, but originally appearing in
Piotto et al.\ 1999). NGC~6441 may host BH stars, although the case
has been unclear (Busso et al.\ 2007; Dieball et al.\ 2009).

To extend the investigations enabled by UV observations of $\omega$
Cen and NGC~2808, we have obtained far-UV and near-UV images of those
additional clusters with the strongest evidence for BH populations:
NGC~2419, NGC~6273 (M19), NGC~6715 (M54), NGC~6388, and NGC~6441.  A
characterization of BH populations in globular clusters can only be
performed definitively with an ultraviolet CMD, where, due to the
small bolometric correction, the photometric effects of enhanced
helium and carbon abundances can be distinguished from changes in
temperature.  These clusters span a wide range of metallicity ($-2.1
\leq $~[Fe/H]~$\leq -0.5$), which enables an exploration of how the
BH population is distributed in temperature, luminosity, and
size, relative to the canonical EHB population, as a function of
cluster environment and metallicity.

To avoid confusion, it is worth clarifying our terminology.  BH stars
were first defined observationally as being hotter and fainter than
the hot end of the canonical EHB.  However, we will show in this paper
that some clusters host stars fainter than the canonical EHB, but with
somewhat redder colors.  These red colors may imply temperatures down
to $\sim$15,000K. The stars appear distinct from the still redder blue
straggler sequence (BSS), and they appear to be associated with the
bluer BH stars.  We will also consider these to be BH stars,
without necessarily implying that they have the same physical origin.

In this paper, we compare our UV CMDs of five massive globular clusters,
together with the previous UV CMD of NGC~2808, to
the expectations from canonical evolution theory and flash-mixing.  In
each case, we explore the expectations for stars born at standard
cluster abundances and for stars born with enhanced helium abundances.
To ensure a consistent exploration of these scenarios, our
evolutionary models are tailored to the appropriate abundance
profiles, as are the model atmospheres and synthetic spectra employed
to transform from the theoretical plane to the observable plane. 

\section{Observations and Data Reduction}

The six clusters discussed herein have all been imaged in the far-UV
and near-UV using the {\it Hubble Space Telescope (HST)}.  Due to
variations in the instrumentation available for each observation, the
bandpasses were not identical for each cluster (see Figure 2), but the effective
wavelengths in each wavelength region (far-UV or near-UV) are similar,
easing comparison of CMDs constructed from these bandpasses.  In Table
1, we summarize the observations for each cluster. 
We also include NGC~2808, the subject of Paper I, observed with
the Space Telescope Imaging Spectrograph (STIS); each STIS UV channel
has pixels of $0.025 \times 0.025$ arcsec and a field of view of
$25 \times 25$ arcsec.

\begin{table*}
\begin{center}
\caption{Ultraviolet Imaging}
\begin{tabular}{lcccc}
\tableline
Cluster & \multicolumn{2}{c}{far-UV} & \multicolumn{2}{c}{near-UV} \\
        & bandpass & exposure (s) & bandpass & exposure (s) \\
\tableline
NGC2419 & ACS/SBC/F150LP  & 5180 & ACS/HRC/F250W & 5080 \\
NGC6273 & ACS/SBC/F150LP  & 1980 & WFPC2/WF3/F255W & 11000 \\
NGC6715 & ACS/SBC/F150LP  & 1980 & WFPC2/WF3/F255W & 11000 \\
NGC2808 & STIS/FUV/F25QTZ & 8916 & STIS/NUV/F25CN270 & 8906 \\
NGC6388 & ACS/SBC/F150LP  & 1980 & WFPC2/WF3/F255W & 11000 \\
NGC6441 & ACS/SBC/F150LP  & 4680 & WFPC2/WF3/F255W & 11000 \\
\tableline
\end{tabular}
\end{center}
\end{table*}

As originally proposed, the observations in the current program intended to 
use the Solar Blind Channel (SBC) and High Resolution Channel (HRC) of the
Advanced Camera for Surveys (ACS) on {\it HST}, with the SBC
F150LP filter providing the far-UV bandpass and the HRC F250W filter
providing the near-UV bandpass.  The ACS/SBC has pixels of $0.034 \times 0.030$
arcsec and a field of view (FOV) of $34.6 \times 30.1$ arcsec,
while the ACS/HRC has pixels of $0.028 \times 0.025$ arcsec and a FOV
of $29 \times 26$ arcsec.  Due to its larger distance and
smaller angular size, the observations of NGC~2419 were planned with a
single tile imaged with the SBC and HRC, while the remaining four
clusters were planned with a $2 \times 2$ mosaic of tiles in each
bandpass.  Due to guide star problems and instrument failures, the
program did not proceed according to plan.

In December 2006, the first observations of NGC~2419 were attempted.
The HRC exposures were obtained in two consecutive orbits, with no
problems.  The SBC exposures were obtained in the subsequent two
orbits, but with serious problems.  The guide star re-acquisition failed in
the first SBC orbit, setting the Take Data Flag down and resulting in
two blank images.  In the final orbit, the re-acquisition of guide
stars was successful, but due to a flight software bug, the telescope
did not complete the maneuvers needed to point at the correct field.
As a consequence, the next two exposures were not blank but offset by
nearly an arcmin from the successful HRC images.  This flight software
bug, which was apparently unique to the {\it HST} two-gyro mode in the
situation where successful guide star acquisitions followed guide star
failures in the same visit, has since been fixed.  The SBC exposures
for NGC~2419 were eventually obtained a year later.  In short,
NGC~2419 was observed with the intended ACS/SBC/F150LP and
ACS/HRC/F250W bandpasses, after some difficulties.

Due to a failure of an ACS power supply in January 2007, the HRC channel
stopped working.  At that time, only the near-UV (HRC) images of NGC~2419
had been taken, so the near-UV images of the remaining four clusters
were obtained with the Wide Field and Planetary Camera 2 (WFPC2),
using the F255W filter and the WF3 chip.  The WF3 chip has $0.1 \times 0.1$
arcsec pixels and a FOV of $80\times 80$ arcsec.  The near-UV channel on WFPC2
is not as sensitive as that on the ACS/HRC, but the larger field of view
allowed imaging of the intended area in a single pointing instead of the
$2 \times 2$ mosaic of HRC images.  Thus, NGC~6273, NGC~6388, NGC~6441,
and NGC~6715 were observed with the ACS/SBC/F150LP and WFPC2/F255W
bandpasses.  

\begin{figure}[t]
\epsscale{1.1}
\plotone{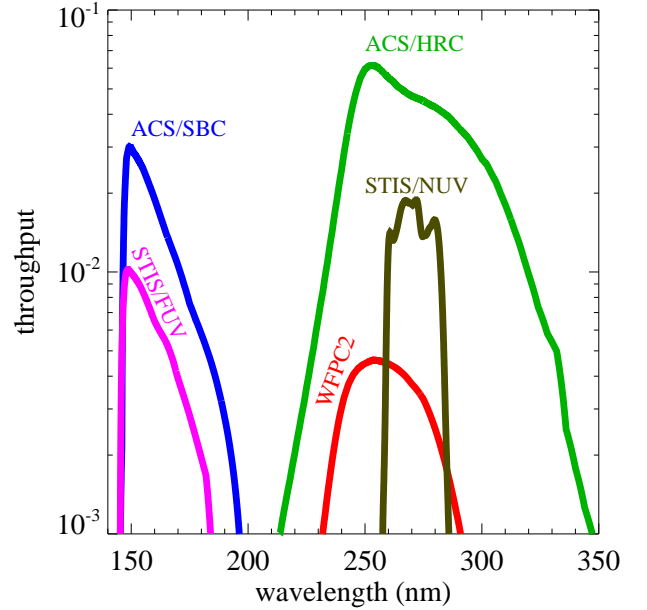}
\epsscale{1.1}
\caption{
The bandpasses employed in our UV imaging (see Table 1).}
\end{figure}

Due to these distinctions, the observations of NGC~2419 and the
remaining four clusters were processed somewhat differently.  For
NGC~2419, the images in each channel were registered and drizzled to a
common scale of 0.$^{\prime \prime}$025, while for the other four
clusters, the scale was 0.$^{\prime \prime}$0455.  The ACS/SBC images
were combined using the drizzle package (Fruchter \& Hook 2002), but
required no cosmic-ray rejection because the SBC is a photon-counting
multianode microchannel array (MAMA).  The WFPC2 and ACS/HRC detectors
are CCDs, so images from these detectors were combined with masking of 
cosmic rays.  

Because globular clusters are much sparser in the UV than in the
optical, we did not perform photometry using fitting of the point
spread function.  Instead, we performed aperture photometry in each
channel with an aperture correction to an infinite aperture, using
the standard aperture corrections for these instruments
(Heyer et al.\ 2004; Maybhate et al.\ 2010; Sirianni et al.\ 2005).  For the
four clusters observed with the ACS/SBC and WFPC2, the apertures were
2.5 pixels and 5 pixels, respectively; for the cluster observed with
the ACS/SBC and ACS/HRC (NGC~2419), the apertures were 2.5 pixels and 1.5
pixels, respectively.  Because it is a MAMA, the SBC requires no
correction for charge transfer inefficiency (CTI), but the near-UV
photometry was corrected for CTI using the corrections of Dolphin (2000)
and Chiaberge et al.\ (2009), for the WFPC2 and ACS/HRC,
respectively.  

The CMD for each cluster is shown in Figure 3.  In Figure 4, we
highlight one of these clusters (NGC~6715) to provide a guide to the
various classes of stars discussed in this paper: blue HB (BHB) stars,
EHB stars, BH stars, BSS, and the
AGB-Manqu$\acute{\rm e}$ (AGBM) progeny of the EHB stars.  Note that
we chose NGC~6715 not because it is particularly representative of 
massive clusters but because each class of star is well-populated in this
CMD.  Each cluster
exhibits a distinct hook feature at the blue end of its HB locus.  As
discussed extensively in Paper I, this hook feature cannot be due to
photometric scatter, differential reddening, or instrumental effects,
because these mechanisms could not affect the photometry of the EHB
without also affecting the photometry of the neighboring BHB stars,
which are only slightly redder; the EHB and BHB stars have similar
far-UV luminosity and photometric errors.  Despite the similar
photometric errors in the EHB and BHB stars, the EHB locus spans a
luminosity range that is several times larger than that on the hot end
of the BHB.

\begin{figure*}[h]
\epsscale{1.1}
\plotone{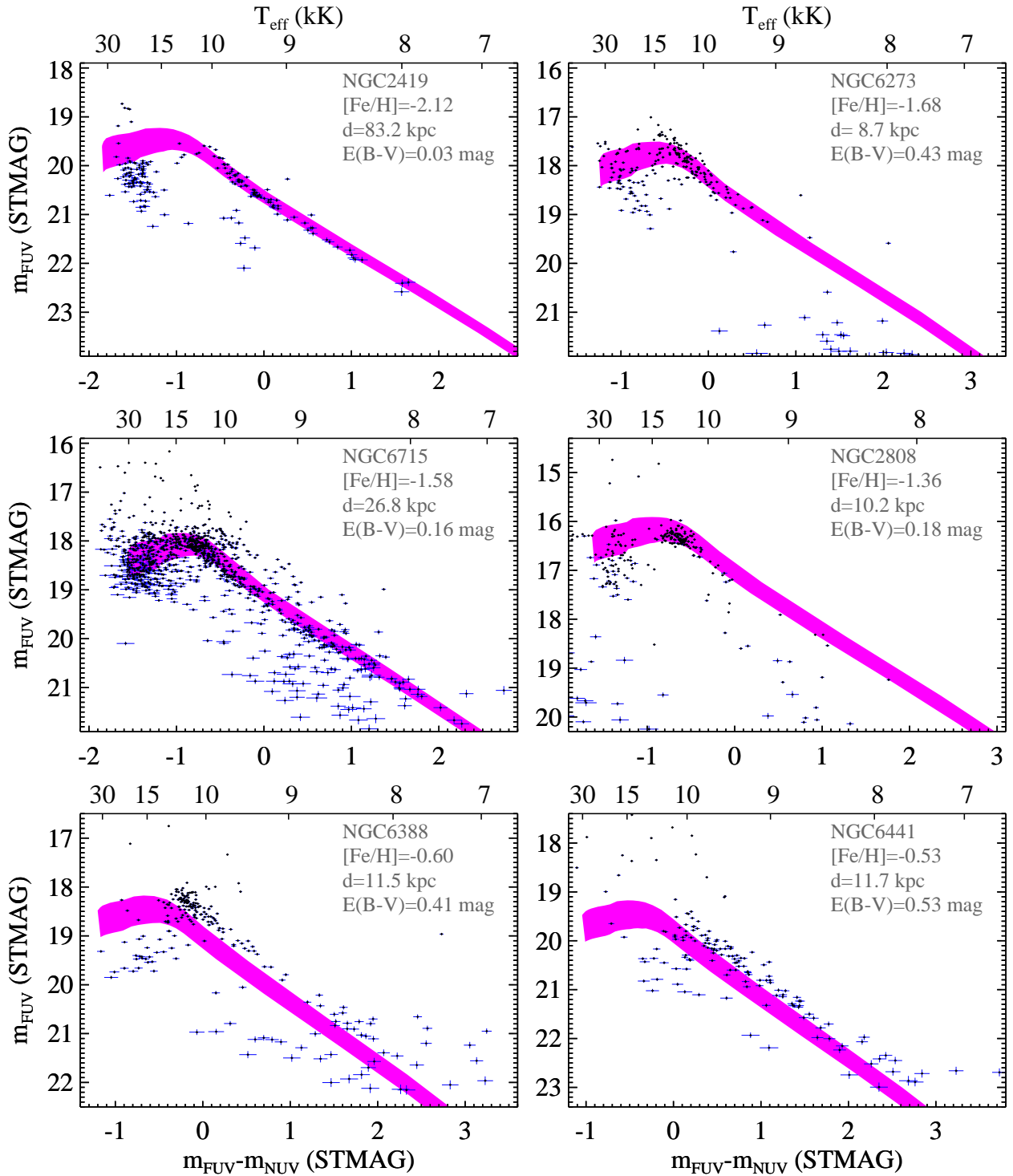}
\epsscale{1.1}
\caption{
Ultraviolet CMDs for six massive globular clusters spanning a wide
range in [Fe/H] ({\it points}), along with the HB locus expected
from canonical stellar evolution theory at standard
helium abundance ($Y=0.23$; {\it violet shaded area}).  The assumed
cluster parameters are indicated ({\it grey labels}).
An approximate conversion (assuming normal cluster abundances) 
between observed color and effective temperature
is shown on the upper abscissa in each panel.  Photometric errors
({\it blue bars}) are only significant for stars much redder and fainter
than the EHB stars that are the focus of this paper.
The BH population appears as a downward hook at the blue
end of the observed HB, where it deviates below the canonical EHB. 
For the two metal-rich clusters, that deviation occurs far to the red
of the hot end of the canonical HB.}
\end{figure*}

\begin{figure}[h]
\epsscale{1.1}
\plotone{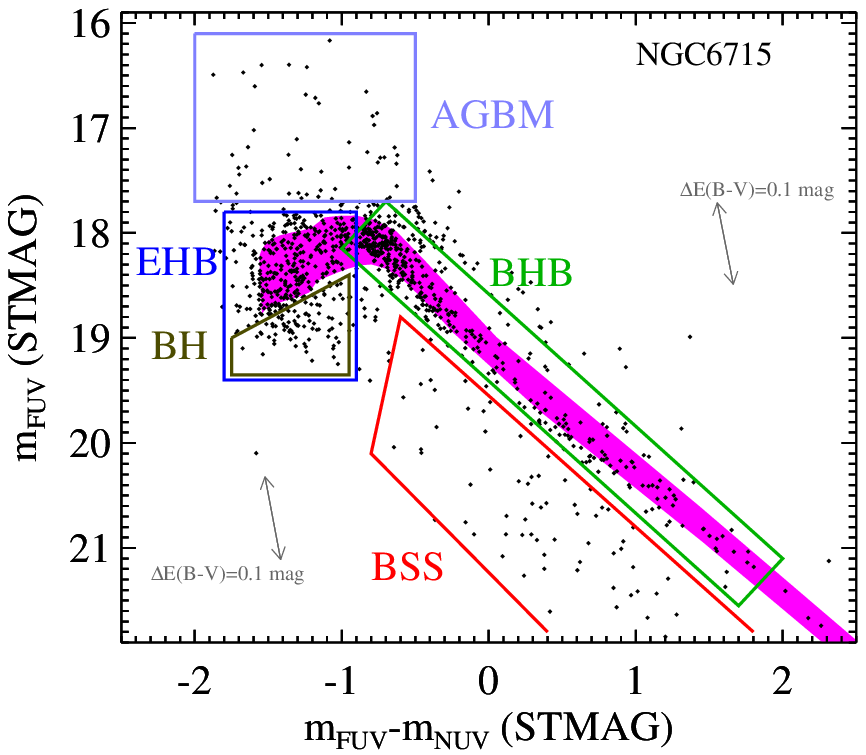}
\epsscale{1.1}
\caption{ The ultraviolet CMD of NGC~6715 ({\it points}), along with
the expected HB locus based upon canonical evolutionary models at
standard helium abundance ({\it violet shaded area}).  The various
classes of stars are indicated ({\it colored boxes}).
The BH stars are the subluminous subset of the EHB stars.  Reddening
vectors for both redder and bluer stars are indicated ({\it labeled
grey arrows})}.
\end{figure}

\section{Models}

\subsection{Stellar Evolutionary Models}

The stellar evolutionary models
used in the present study were computed with
a procedure similar to the one described in Paper I.  For
each cluster, we constructed grids of evolutionary sequences that followed
the evolution of appropriate low-mass stars continuously from the zero-age
MS through the end of the HB phase for two values of the initial
helium abundance: a helium-normal abundance $Y$ of 0.23, representing the
helium abundance in stars born during the first stellar generation with
standard cluster abundances, and a helium-rich abundance $Y$ of 0.40,
representing a reasonable upper limit on the helium enhancement in the
second stellar generation born after a process of cluster self-enrichment.
The hotter HB sequences in these grids were also evolved through the
post-HB phases.

The heavy element abundance $Z$ for these sequences was determined
from the observed [Fe/H] value for each cluster, assuming
an $\alpha$-element enhancement of [$\alpha$/Fe]~=~0.3, as appropriate
for metal-poor globular cluster stars (Carney 1996).  Using the
prescription given by Salaris et al.\ (1993) for converting
an $\alpha$-enhanced abundance into an equivalent scaled-solar
abundance, we obtained scaled-solar heavy element abundances $Z$ of
0.00023, 0.00064, 0.00081, 0.0015, 0.0077, and 0.0090, respectively,
for the six clusters NGC~2419, NGC~6273, NGC~6715, NGC~2808, NGC~6388
and NGC~6441.  We used the same $Z$ value to compute both the helium-normal
and helium-rich sequences for each cluster.  The initial main sequence
mass for each composition ($Y$, $Z$) was calculated for an age of 13
Gyr at the tip of the RGB, although ages a few Gyr younger would
have little qualitative effect on the subsequent evolutionary behavior.

The sequences for each composition ($Y$, $Z$) differ from each other
only in the amount of mass loss along the RGB, which we parameterize
with the Reimers (1975, 1977) mass-loss formulation:

$\dot{M} = -4 \times 10^{-13} \eta_R L / g R $~~~~($M_\odot$ yr$^{-1}$),

\noindent
where $L$, $g$, and $R$ are the stellar luminosity, gravity, and
radius, respectively, in solar units, and $\eta_R$ is the well-known
Reimers mass-loss parameter.  Because our calculations follow the
evolution through the helium-core flash, we are able to determine how
the evolutionary path from the MS to the ZAHB depends on the assumed
mass-loss rate (see Figure 1).  The red end of the ZAHB is set by the
sequence with $\eta_R$ = 0.0 (no mass loss).  Up to a given
value of $\eta_R$ ($\sim$0.7 for intermediate [Fe/H] values), the
models remain tightly bound to the RGB until the helium-core flash
(Figure 1a).  However, for higher values of $\eta_R$, the
models leave the RGB and evolve to high effective temperatures before
igniting helium (Castellani \& Castellani 1993), either during the
crossing of the CMD (Figure 1b) or while
descending the WD cooling curve (Figure 1c).
D'Cruz et al.\ (1996) suggested that such ``hot flashers'' might
provide another channel for populating the EHB.  Although the masses
of the late hot flashers for each composition ($Y$, $Z$) fall within a
narrow range, they are produced over a rather wide range in the mass
loss rate ($\Delta \eta_R$~$\sim$0.1).  Extremely fine tuning of the
mass-loss rate is thus not required to produce a late hot flasher  
(see Paper I).  For sufficiently high values of $\eta_R$, the models
simply cross the HR diagram and descend the WD cooling curve
without igniting helium, thus dying as helium WDs (see Figure 1d).

The RGB tip flashers and early hot flashers ignite helium at a time
when the models are very luminous.  The strong hydrogen-burning shell
in these models forms a high entropy barrier that prevents the flash
convection from penetrating into the envelope, and hence no mixing
then occurs between the envelope and the core.  Thus any star that
undergoes the helium-core flash before reaching the WD cooling curve
will arrive on the ZAHB via canonical evolution (i.e., with a normal
atmospheric composition).  In contrast, stars that ignite helium on the
WD cooling curve have much weaker hydrogen-burning shells and thus
much lower entropy barriers (Sweigart 1997; Paper I).  Under these
conditions, penetration of the flash convection into the envelope
becomes inevitable.  This flash mixing will mix hydrogen-rich material
from the envelope into the core and helium- and carbon-rich material
from the core into the envelope, thereby greatly enhancing the surface
helium and carbon abundances.  A similar process may produce other
extremely hydrogen-poor stars (also known as ``born again'' stars) during a
very late helium-{\it shell} flash (Renzini 1990).

Due to the numerical
difficulties associated with this mixing process (Paper I), we
terminated our sequences as soon as they encountered flash mixing.
Nevertheless, we can still predict the surface composition that these
flash-mixed models should have when they arrive on the ZAHB.  The
available calculations clearly indicate that flash mixing will consume
most, if not all, of the envelope hydrogen.  Thus we expect the
surface composition of the flash-mixed models to be strongly depleted
in hydrogen and enriched in helium.  The surface carbon abundance
should also be close to its value in the flash convection zone,
namely, 4\% by mass.  For these reasons the flash-mixed models given
in this paper will assume that the surface carbon abundance has been
increased to 4\% by mass and that all of the remaining surface
hydrogen has been burned to helium.  It is worth noting that Cassisi
et al.\ (2003) and Miller Bertolami et al.\ (2008) were able to follow
the flash mixing process in detail, and their values for the
flash-mixed surface composition are in good agreement with the
composition adopted here.  Note also 
that some of the carbon may be burned to nitrogen
during the flash mixing, depending on the details of the nucleosynthesis;
such an outcome would
not have a significant effect on the stellar evolution models.

\subsection{Non-LTE Model Atmospheres and Synthetic Spectra}

In order to produce theoretical spectra for the evolutionary
tracks, we calculated a series of 60 non-LTE line-blanketed model
atmospheres with our model atmosphere program, {\sc Tlusty\/}, version
201\footnote{Available at http://nova.astro.umd.edu}.  For each
globular cluster, we used the appropriate metallicities and helium
abundances for the various evolutionary scenarios explored.  
{\sc Tlusty\/} computes stellar model photospheres in a plane-parallel
geometry, assuming radiative and hydrostatic equilibria.  Departures
from LTE are explicitly allowed for a large set of chemical species
and arbitrarily complex model atoms, using our hybrid Complete
Linearization/Accelerated Lambda Iteration method (Hubeny \& Lanz
1995).  More specifically, the model atmospheres allow for departures from LTE
for 1132 levels and superlevels of 52 ions, \ion{H}{1}, \ion{He}{1},
\ion{He}{2}, \ion{C}{1} -- \ion{C}{4}, \ion{N}{1} -- \ion{N}{5},
\ion{O}{1} -- \ion{O}{6}, \ion{Ne}{1} -- \ion{Ne}{4}, \ion{Mg}{2},
\ion{Al}{2}, \ion{Al}{3}, \ion{Si}{2} -- \ion{Si}{4}, \ion{P}{4},
\ion{P}{5}, \ion{S}{2} -- \ion{S}{6}, \ion{Fe}{2} -- \ion{Fe}{6}.
Details of the model atom setup are provided in Lanz \& Hubeny (2003,
2007), and in Cunha et al.\ (2006) for updated Neon models.

The model atmospheres span the range of parameters suitable
for most EHB stars: 20,000~K$\leq$~$T_{\rm eff}$~$\leq$50,000K (with a
5,000 K step), 4.5~$\leq$~log~$g$~$\leq$~6.0 (0.5 dex step), and
$V_{turb}$~=~5~km/s. We assumed three different helium compositions:
standard helium ($Y$~=~0.23), helium-rich ($Y$~=~0.4), and
flash-mixed ($Y$~=~0.96, with carbon and nitrogen mass
fractions of 3\% and 1\%, respectively; Lanz et al.\ 2004 -- hereafter
Paper II).  Note that the assumption of 4\% carbon or 4\% nitrogen (or
some intermediate mix) makes little difference in broad-band UV colors
(see Paper I). 

With our spectrum synthesis code, {\sc Synspec\/}, we then calculated
the detailed emergent ultraviolet spectrum of each model atmosphere.
Quadratic interpolations in the spectra database were performed to
derive theoretical far-UV and near-UV magnitudes for the specific stellar parameters
along each evolutionary track.  The parameters of the quadratic interpolation
are wavelength-dependent. Detailed tests show that this interpolation
process yields colors that do not differ by more than 0.01 mag,
compared to using a model atmosphere calculated with the exact stellar
parameters. This approach saves significant computing time without
significantly affecting the final accuracy.

\subsection{Transfer of Models to the Color-Magnitude Diagrams}

With these models in hand, we transformed the evolutionary models for
each cluster to the observable plane.  For each cluster, we
interpolated in the grid of synthetic spectra having the appropriate
abundance pattern (cluster metallicity with $Y=0.23$, cluster
metallicity with $Y=0.4$, or the flash-mixed composition).  For stars
cooler than the lowest effective temperature in our non-LTE grids, we
interpolated in the Castelli \& Kurucz (2003) spectral grids at enhanced
[$\alpha$/Fe].  The interpolated spectrum for each stellar
evolutionary model was reddened using the average Galactic extinction
curve of Fitzpatrick (1999), and then passed through the relevant
near-UV and far-UV bandpasses using the IRAF {\sc synphot} package, resulting
in magnitudes in the STMAG system, where $m= -2.5 \times $~log$_{10}
f_\lambda -21.1$~mag.  Note that this method appropriately applies 
the reddening in a bandpass-dependent and $T_{\rm eff}$-dependent manner. 

\section{Comparison of Photometry to Models}

\subsection{Canonical Models at Standard Helium Abundance}

We first compare
our CMDs to the expectations from canonical stellar evolutionary models
with a standard helium abundance.  We use the BHB as a
standard candle to guide the normalization of these models.
Specifically, we adopt metallicity and distance values that are
representative of values in the literature, and then make small
adjustments to the extinction so as to place the BHB of a canonical
model at the expected location.  Note that the reddening vector in our
$m_{FUV}$ vs. $m_{FUV} - m_{NUV}$ CMDs is nearly vertical, so one
could adjust the distance instead of the reddening, or some
combination thereof; a change of 0.01~mag in $E(B-V)$ is equivalent to
a change of 0.079~mag in distance modulus.  For the four metal-poor
clusters, the theoretical HB distribution should be coincident with
the observed HB locus at the BHB.  For the two metal-rich clusters, Rich et
al.\ (1997) have shown that the HB slopes strongly upward as one moves
from the reddest HB stars to the BHB (see also Busso et al.\ 2007),
such that the BHB is $\sim$0.5~mag brighter in the optical
than one would expect for
a canonical locus normalized to the red HB.  It would be simpler to 
consistently align our models to the red HB in all of our clusters, 
but it falls outside of our UV CMDs, driving our use of the BHB;
thus, for the two metal-rich clusters we choose an extinction that
places the canonical locus 0.5~mag fainter in the far-UV
than the observed BHB locus.  We are assuming that a star at a given reference
point on the BHB (defined at a particular effective temperature) has changed 
in bolometric luminosity, such that the offset in the far-UV is the same 
0.5~mag observed in the optical.  The physical motivation for this assumption 
comes from the fact that the bolometric luminosities of all HB stars in the 
range $6000 \lesssim T_{\rm eff} \lesssim 18000$~K are increased by nearly
the same amount when $Y$ is enhanced, and that the effect on the 
UV-to-optical color in the emergent spectrum is minimal.
As we shall see in the next section, our choice of extinction
is equivalent to aligning canonical BHB models of
intermediate $Y$ ($\sim$0.3) to the observed BHB locus.

In Figure 3, we show the CMD of each cluster compared to the expected
locus for stars in the core helium-burning stage, assuming standard
cluster metallicity and helium abundance.  The comparison is made over a
wide color range that includes stars much redder than the EHB.  In
Figure 5, we restrict the comparison to those stars hotter than
$\sim$9,000~K, and shade the HB locus to indicate the timescales
expected for HB evolution.  We expect most of the HB stars to follow close to
the low-luminosity edge of the canonical helium-burning locus, because the
evolution accelerates towards the latter stages of the HB phase
(i.e., from the bottom to the top of
the canonical HB locus).  Because we use the BHB as a standard candle
to normalize the models, we can make an appropriate comparison of the observed
EHB locus and the predicted EHB locus.

\begin{figure*}[h]
\epsscale{1.1}
\plotone{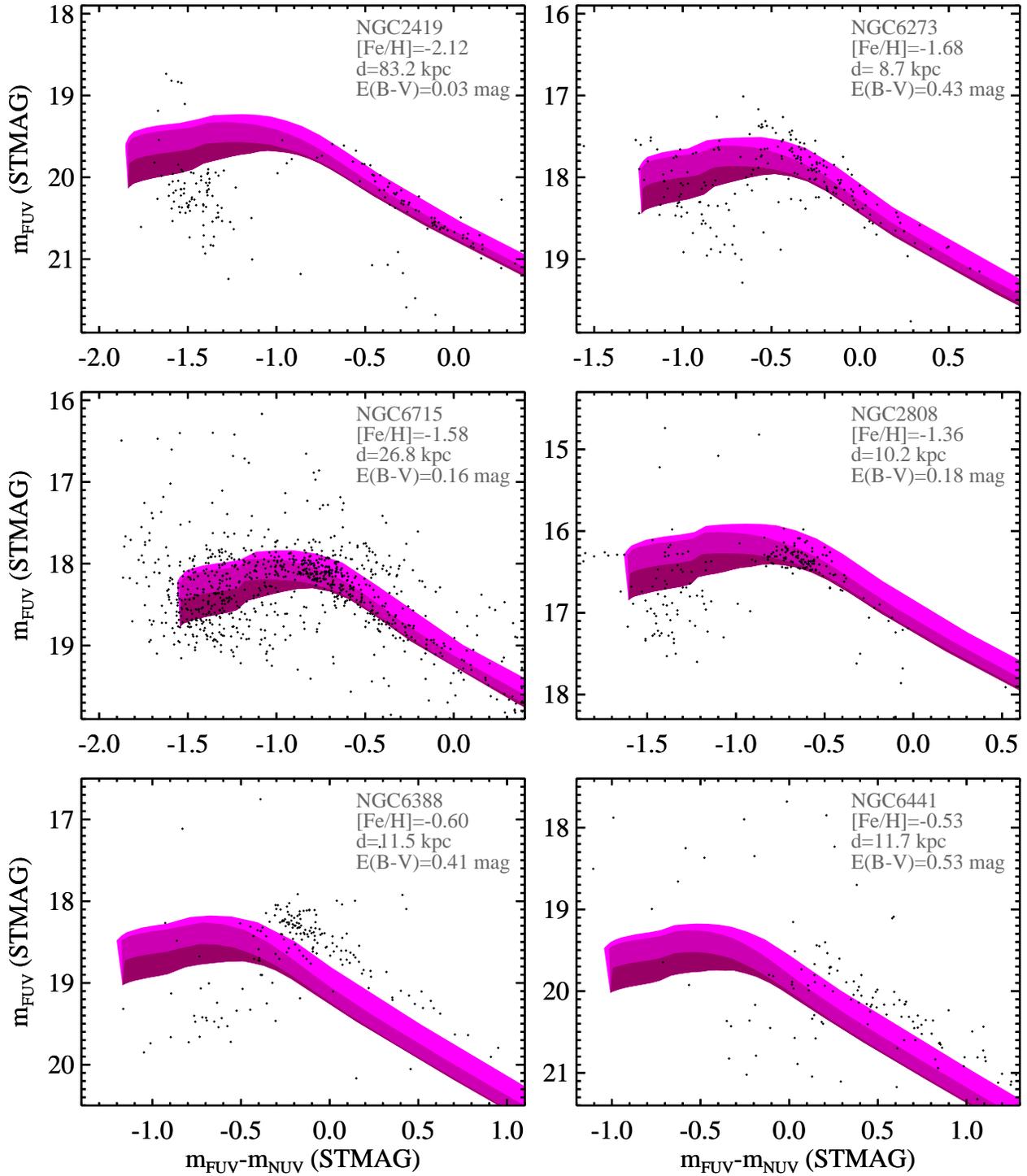}
\epsscale{1.1}
\caption{
The ultraviolet CMDs shown in Figure 3, but restricted to the brighter and
bluer stars in order to highlight the hot end of the HB.
The canonical HB locus for each cluster is shown with
shading to indicate evolutionary timescales; the three levels
of shading, from darkest to lightest, indicate zero-age to 50\% HB lifetime,
50\% to 90\% HB lifetime, and 90\% to 99\% HB lifetime.}
\end{figure*}

{\bf NGC~2419} We have assumed a metallicity of [Fe/H]~$= -2.12$
(Harris 1996) and a distance of 83.2 kpc (Ripepi et al.\ 2007).  For
the reddening, our assumption of $E(B-V) = 0.03$~mag is somewhat lower
than that assumed elsewhere (0.08--0.11~mag; Harris 1996; Piotto et
al.\ 2002; Ripepi et al.\ 2007), but the upward shift 
in the models is required to align the observed and theoretical loci
at the BHB.  As one progresses blueward in $m_{FUV} - m_{NUV}$ color
from 1.0 to $-0.8$~mag, the observed HB appears to slope slightly
upward relative to canonical HB, but it is worth noting that optical
CMDs spanning the $B$, $V$, and $I$ bandpasses (Piotto et al.\ 2002;
Ripepi et al.\ 2007) do not show a sloping HB that would justify
normalizing the canonical locus significantly below the BHB.  As one
moves to stars bluer than the BHB, the observed HB locus drops
abruptly to fainter magnitudes, such that nearly all of the observed
EHB stars are subluminous (i.e., BH stars).

{\bf NGC~6273} We have assumed a metallicity of [Fe/H]~$=-1.68$ and a
distance of 8.7~kpc (Piotto et al.\ 2002).  For the reddening, our
assumption of $E(B-V) = 0.43$~mag is larger than that assumed by
Piotto et al.\ (2002; 0.37~mag), shifting the theoretical locus fainter by
$\sim$0.5~mag in the far-UV to give alignment with the observed BHB.  As with
NGC~2419, the HB slope appears somewhat steeper than expected as one
moves blueward from 0.5~mag to $-0.5$ mag in $m_{FUV} - m_{NUV}$
color, even though the optical CMD (e.g., Piotto et al.\ 2002) shows no
indication of an upward HB slope from the red HB to the BHB.  At the
hot end of the HB, the observed population is roughly split between
normal EHB stars and BH stars.

{\bf NGC~6715} We have assumed a metallicity of [Fe/H]~$=-1.58$ and a
distance of 26.8 kpc (Harris 1996).  For the reddening, our assumption
of $E(B-V) = 0.16$~mag is slightly higher than that assumed elsewhere
(0.15~mag; Harris 1996; Siegel et al.\ 2007), shifting the models
fainter by 0.08~mag to align with the observed BHB.  The observed HB
locus and the theoretical locus agree well over the entire BHB range,
but at the EHB, the observed population is roughly split between
normal EHB stars and BH stars.  Note that NGC~6715 may actually
represent the core of the disrupted Sgr dwarf galaxy 
(e.g., Layden \& Sarajedini 2000; Siegel et al.\ 2007).  Even if it is not 
the core of the Sgr dwarf, the earlier optical CMDs in this vicinity exhibit 
populations more complex than one would usually associate with a
globular cluster, which may explain the large scatter in our UV CMD.

{\bf NGC~2808} This cluster was examined extensively in Paper I.  The
cluster is included here for comparison, and we assume the same
cluster parameters used previously.  Although the HB sample in
NGC~2808 is smaller than that in NGC~6715, the morphology of the
observed HB locus is similar.  The similarities between the NGC~6715
and NGC~2808 CMDs may be related to the similarities in metallicity
and central concentration; NGC~6273 is also close in metallicity, but
is less centrally concentrated and less luminous than NGC~6715 and NGC~2808
(Harris 1996).

{\bf NGC~6388} We have assumed a metallicity of [Fe/H]~$=-0.6$ and a
distance of 11.5~kpc (Piotto et al.\ 2002).  For the reddening, our
assumption of $E(B-V) = 0.41$~mag is slightly larger than that of
Piotto et al.\ (2002; 0.4~mag), placing the canonical locus 0.5~mag
fainter than the observed BHB, as found by Busso et al.\ (2007) and
discussed above.  However, Busso et al.\ (2007) assumed a distance of
11.2~kpc and $E(B-V) = 0.45$~mag; if we adopted these values, the
canonical HB locus would be an additional 0.3~mag fainter than we have
assumed.  In contrast, assuming the shorter distance (10 kpc) and
lower reddening (0.37~mag) of Harris (1996) would align the canonical
locus with the observed HB locus at the BHB, as we did for the
metal-poor clusters.  Although we have chosen an extinction that
places the canonical BHB 0.5~mag below the observed BHB, nearly all of
the observed EHB stars are subluminous with respect to the theoretical locus, as
also seen in NGC~2419.  However, unlike the situation in NGC~2419, the
subluminous EHB stars of NGC~6388 cover a much wider range of
color than that seen in any of the metal-poor clusters; the
subluminous EHB stars in NGC~2419 are well-separated in color from the
BHB stars, but there is no such color gap in the NGC~6388 population.
Despite their extension to unusually red colors, the subluminous EHB
stars in NGC~6388 are much bluer than the BSS (see Figure 3), 
and thus there is no
confusion between these two classes of stars.  Moreover, 
the subluminous EHB stars
in NGC~6388 comprise a contiguous and distinct group, which suggests
that they may all be BH stars, including the reddest stars in this group.

{\bf NGC~6441} We have assumed a metallicity of [Fe/H]~$=-0.53$, and a
distance of 11.7~kpc (Harris 1996). For the reddening, our assumption
of $E(B-V) = 0.53$~mag is larger than that of Harris (1996; 0.47~mag),
placing the canonical locus 0.5~mag fainter than the observed BHB, as
found by Busso et al.\ (2007) and discussed above.  However, assuming
the distance (15.2 kpc) and reddening (0.48~mag) of Busso et
al.\ (2007) would place the canonical locus an additional 0.2~mag
fainter than we have assumed.  As with NGC~6388, assuming the Harris
(1996) distance and reddening would align the canonical locus with the
observed HB locus on the BHB, which is the normalization we used with
the metal-poor clusters.  Piotto et al.\ (2002) assumed a shorter
distance (11.3~kpc) and less reddening (0.44~mag); assuming these
values would place our models above the observed HB locus for all
temperatures (i.e., not simply the potential BH stars).  Regardless of
the normalization, the colors imply that
there are almost no stars hotter than $\sim$15,000~K in the
cluster on or below the canonical EHB.
There are stars below the canonical HB in
the temperature range of 10,000 -- 15,000~K, which is hotter than the
BSS.  When considered in the context of the NGC~6388 BH population
(which extends far to the red), it is possible that these subluminous
stars in NGC~6441 are BH stars, despite their relatively red color.
We inspected archival WFPC2 images in the F336W ($U$) and F439W ($B$) 
bands to determine where these unusually red BH candidates would 
fall in optical CMDs.  Although the core of the cluster is crowded at 
these longer wavelengths, five such stars were sufficiently isolated to
obtain clean photometry, and in the $B$ vs. $U-B$ CMD they clearly fall
within the low-luminosity end of the long blue HB tail, reinforcing the 
need for UV photometry to distinguish between temperature and luminosity 
variations in hot stars.

In all six clusters, the hottest HB stars observed are subluminous
with respect to canonical models (see Figures 3 and 5).  Thus there
are BH stars in each cluster, with the possible exception
of NGC~6441, where these subluminous stars are all relatively red.
It is clear that canonical models at $Y=0.23$ cannot explain the HB populations
of these clusters.  Regardless of how such models are normalized
via varying the assumed reddening or distance,
they cannot reproduce the observed luminosity difference between
the BHB and the subluminous EHB stars.

\begin{figure*}[t]
\epsscale{1.1}
\plotone{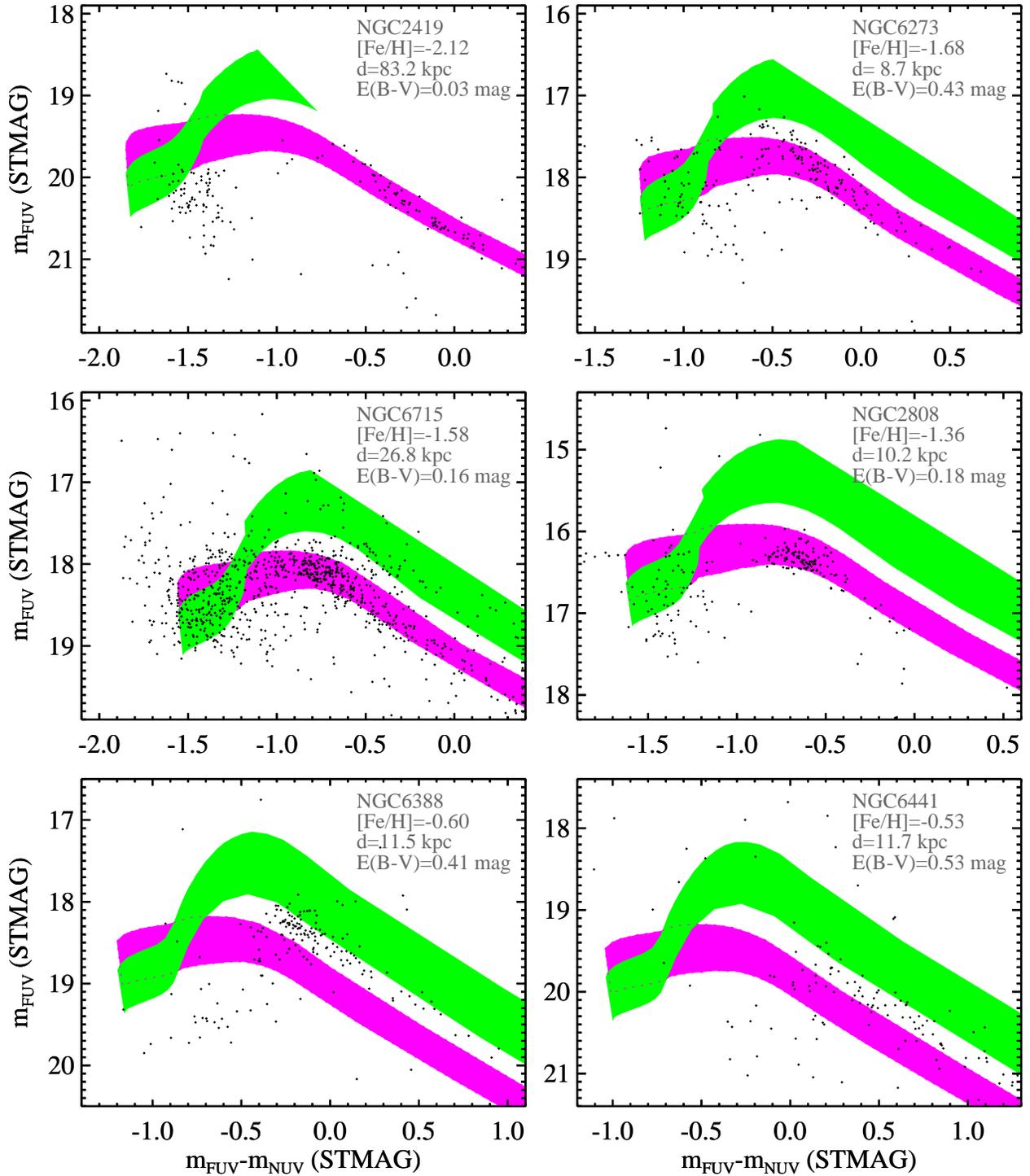}
\epsscale{1.1}
\caption{The same as in Figure 5, but with the canonical locus at standard
helium abundance ($Y=0.23$; {\it violet}) compared to the canonical locus with
enhanced helium ($Y=0.4$; {\it green}).  The boundaries of the partly-obscured
$Y=0.23$ locus are indicated for clarity ({\it violet dashed lines}).} 
\end{figure*}

\subsection{Canonical Models with Enhanced Helium Abundance}

We next consider models with the cluster metallicity
but with the helium abundance increased to $Y=0.4$.
This value of helium enhancement is at the upper end of the plausible
range for stars that have been enhanced at birth due to cluster
self-enrichment.  In Figure 6 we compare
these models with $Y=0.4$ to those with $Y=0.23$,
assuming the same cluster parameters ([Fe/H], 
distance, and reddening) used previously.

Figure 6 shows that an enhanced helium abundance produces a more pronounced
dropoff in luminosity between the BHB and
the EHB.  Relative to the $Y=0.23$ locus, the
$Y=0.4$ locus is fainter on the EHB but much brighter on the BHB.  The
fainter EHB luminosity of the $Y=0.4$ locus is due to a slightly smaller
core mass while the brighter BHB luminosity is due to a greater energy
output of the H-burning shell.  This dependence of the luminosity
difference between the BHB and EHB on the helium abundance is well
established and was evident even in the early HB calculations of
Sweigart \& Gross (1976).  We note in Figure 6 that the $Y=0.4$ models
for the very metal-poor cluster NGC~2419 are restricted to hot
temperatures, even for those HB stars that lost no mass on the RGB.
This is because the main sequence turnoff mass for a population at
$Y=0.4$, [Fe/H]=$-2.12$, and an age of 13 Gyr is only 0.614~$M_\odot$; this 
low turnoff mass, combined with this metal-poor composition, will only
produce blue HB stars, even with no mass loss.

It is
clear that the $Y=0.4$ models do not agree well with the observed
photometry when we assume the same parameters ([Fe/H], distance, and
reddening) used in the $Y=0.23$ models.  The hot end of the $Y=0.4$
locus is not as faint nor as red as the observed BH stars, and the
cooler part of the $Y=0.4$ locus is much brighter than the observed
BHB stars.
While a combination of the $Y=0.23$ and $Y=0.4$ models
would lead to a larger luminosity width for the EHB,
the predicted range in luminosity would still not extend to the faint
magnitudes reached by the BH stars.  It is also clear that any range
in $Y$ along the BHB must be fairly small, given the narrow luminosity
width of the BHB loci in Figure 6.  Note that the larger scatter in NGC~6715
may be due to the complex populations associated with the Sgr dwarf galaxy
(Layden \& Sarajedini 2000; Siegel et al.\ 2007).

At first glance, one might think that the agreement between
$Y=0.4$ locus and the observed BHB and EHB stars could be improved
by assuming a larger reddening or distance,
because such adjustments to the cluster parameters would shift
the models to fainter magnitudes.  Indeed, if the $Y=0.4$ locus
were shifted until it fit the observed BHB, the lower edge of the
EHB for these models would be in much better agreement with the
observed lower edge of the BH stars.  However, such adjustments
to the reddening or distance would violate important observational
constraints.

If we attempt to fit the BHB with the $Y=0.4$ locus, we will have
to confront the following observational difficulties.  Optical CMDs of
NGC~2419 (Sandquist \& Hess 2008), NGC~6273 (Piotto et al.\ 1999),
NGC~6715 (Layden \& Sarajedini 2000), or NGC~2808 (Bedin et al.\ 2000)
do not show a significant difference in the $V$-band luminosity of the red HB
and BHB stars, implying that any difference in the helium abundance
between these stars must be small.  Thus the cluster parameters
we derived in fitting the BHBs of these clusters to the canonical $Y=0.23$
locus should also apply to the red HB stars.  In the case of the two
metal-rich clusters NGC~6388 and NGC~6441, 
it is well known that the HB slopes upward (Rich et al.\ 1997; 
Busso et al.\ 2007), with the $V$-band luminosity of the BHB
being $\sim$0.5~mag brighter than that of the red HB clump.  The BHB stars
in these clusters may thus have an intermediate helium
abundance ($Y\sim 0.3$).  However, as discussed above, we took
this offset in the BHB luminosity into account when fitting
the $Y=0.23$ model to our UV CMDs.  Thus our parameters for
these two clusters should also apply to their red HB clumps.  If
the BHB should be dominated by stars with $Y=0.4$ and the red HB
by stars with $Y=0.23$, one would therefore predict a luminosity
difference between the BHB and red HB clump that is much larger than
observed in any of our six clusters.  To avoid this contradiction,
one might argue alternatively that the red HB stars are themselves helium-rich,
but this would imply a much brighter red HB luminosity and hence
a much younger (and untenable) cluster age from the observed
luminosity difference between the red HB and the MSTO (e.g., see
Mart$\acute{\rm i}$n-Franch et al.\ 2010).
As one example, Catelan et al.\ (2006) have compared the CMDs
of NGC~6388 and 47~Tuc and concluded that both of these clusters
have similarly old ages.

In short, while the $Y=0.4$ locus exhibits a sharp drop in luminosity
between the BHB and EHB, these models cannot by themselves explain
the BH phenomenon.  Given the lack of helium enhancement on the red HB
and the relative luminosities of the BHB and red HB, we cannot align
models of arbitrarily high $Y$ to the BHB without violating other
observational constraints.  If we assume the BHB
has a standard helium abundance ($Y=0.23$, in the case of the four metal-poor
clusters) or an intermediate helium abundance ($Y\sim 0.3$, in the case of the
two metal-rich clusters), then even a wide range
of helium abundance ($0.23 \lesssim Y \lesssim 0.4$) on the EHB
would not explain either the faint luminosities or the wide color range
observed in the BH stars.

\begin{figure*}[t]
\epsscale{1.1}
\plotone{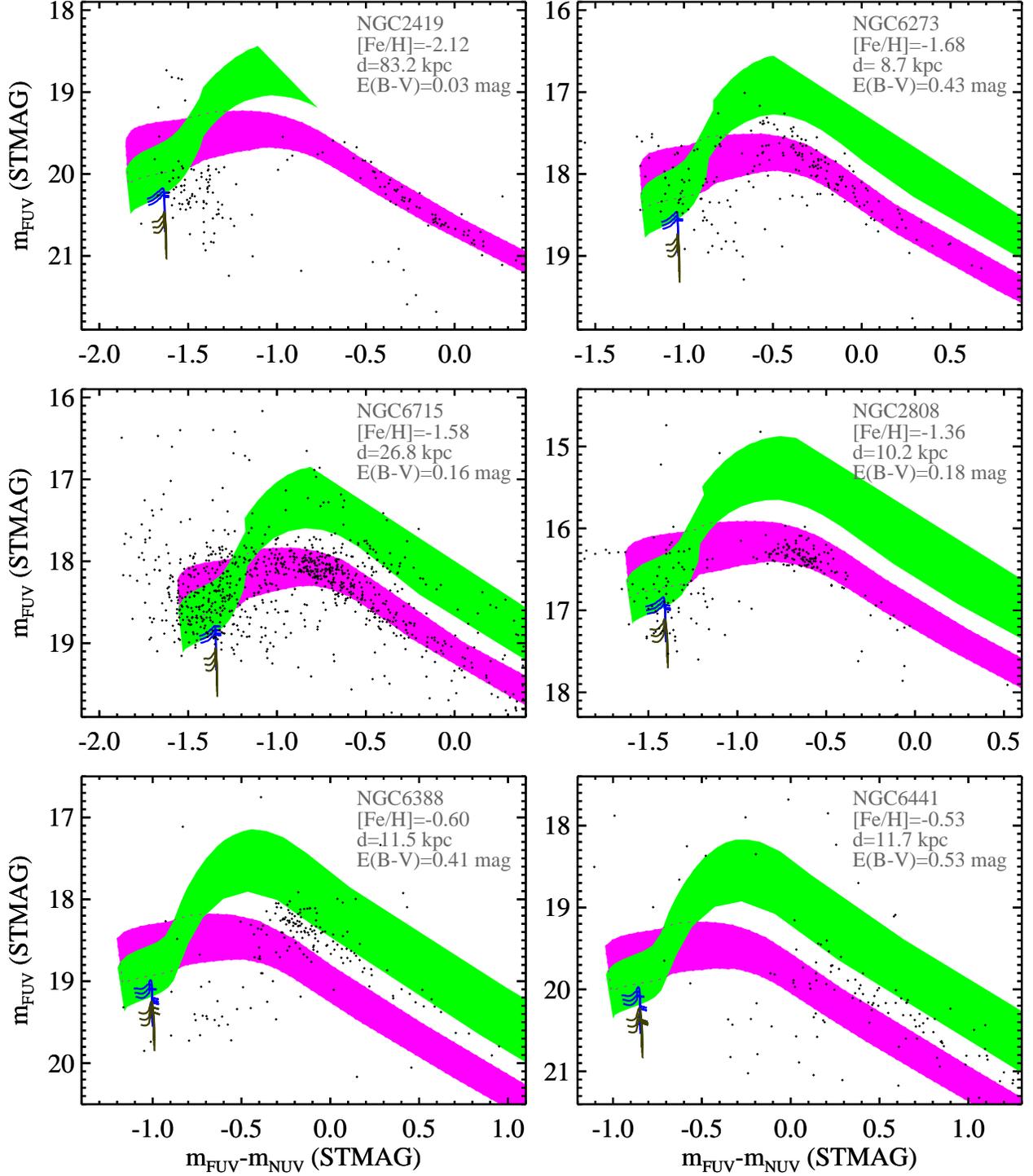}
\epsscale{1.1}
\caption{Our UV CMDs compared to canonical models at standard helium
abundance ($Y=0.23$; {\it violet}), canonical models at enhanced helium
abundance ($Y=0.4$; {\it green}), and flash-mixed models for stars
born at standard helium abundance ({\it blue curves}) and enhanced helium
abundance ({\it brown curves}).  
The combination of canonical and flash-mixed models is able to
reproduce the full luminosity width of the observed EHB, but not the
red colors seen in some of the BH stars.}
\end{figure*}

\subsection{Flash Mixed Models}

We next consider models which undergo flash mixing due to high
mass loss on the RGB.  In Figure 7, we plot these flash-mixed models
in our UV CMDs, assuming the same reddening and distance values obtained 
in section 4.1.  Those values align a canonical $Y=0.23$ locus to the BHB 
in the four metal-poor clusters, but place this $Y=0.23$ locus 0.5~mag fainter 
(in $m_{FUV}$) than the BHB in the two metal-rich
clusters.  For each cluster in Figure 7 we show the HB evolutionary tracks
for three representative flash-mixed models; two of these tracks
fall at the extremes of the range in the mass loss rate leading to flash mixing,
as measured by the value of $\eta_R$, while one falls halfway
between these extremes.  We plot these triplets of
flash-mixed tracks for both $Y=0.23$ and $Y=0.4$
in order to bracket the range of behavior expected
from any sub-population in the cluster that
undergoes enough mass loss for flash mixing.
All of the flash-mixed tracks for each cluster fall within a narrow temperature
range.  The offset in luminosity among these tracks is due
to small differences in the helium-core mass.

As can be seen in Figure 7, flash-mixed models can reproduce the faint
luminosities observed in the BH population, particularly the
flash-mixed models originating in a subpopulation with enhanced helium.
Flash mixing thus offers a way of reproducing the faint luminosities
observed in the BH stars without causing problems at the BHB or red HB
clump, as discussed in the last section.  Furthermore, the flash-mixed models,
when combined with the canonical HB locus, are better able to reproduce the
large luminosity width observed at the hot end of the EHB.  

There is one obvious problem, however, which was also seen in the
previous sections; like canonical models, flash-mixed models
(whether originating in a population with enhanced or standard $Y$)
cannot reproduce the red colors in the BH population.  Paper I
explored various ways of producing such red BH stars,
including enhanced line blanketing in the atmosphere, due to an increase in
the atmospheric Fe abundance via radiative diffusion.  
In the case of NGC~2808, an
increase in the [Fe/H] from $-1.36$ to $+1.0$ could shift the BH stars
$\sim$0.3~mag to the red.  However, this color shift is not enough to
explain the full color range of the BH population.  Furthermore, the
effect would be even smaller for the metal-rich clusters, which are
the very clusters exhibiting the reddest BH stars.  Because the
metal-rich clusters already incur significant line blanketing,
increasing the [Fe/H] via diffusion does not help much; for example,
increasing the [Fe/H] in NGC~6388 from $-0.6$ to $+1.4$ only shifts
the BH color redward by 0.26~mag.  Thus, at this time, the mechanism for
producing subluminous HB stars with such red colors remains a puzzle,
and it suggests that the formation mechanism for these stars is not
yet fully understood.

\section{Discussion}

\subsection{Blue Hook Stars}

In this paper, we have compared the UV CMDs of six massive globular
clusters with the expectations from stellar evolution theory.  We find
that in each of these clusters, the hottest HB stars are subluminous
with respect to canonical models, whether or not those models assume
a standard helium abundance ($Y=0.23$) or an enhanced helium abundance
($Y=0.4$).  The most straightforward interpretation is that all of
these clusters host BH stars.  However, in the two metal-rich
clusters, the subluminous stars either extend to relatively red colors
(NGC~6388) or are restricted to these red colors (NGC~6441), which
might imply that an additional or distinct phenomenon is at work at
high metallicity.  Indeed, the existence of BH stars in NGC~6441 has
been questioned (e.g., Dieball et al.\ 2009).  The BH population of
NGC~6388 may be the link between the BH populations in the metal-poor
clusters and that in NGC~6441, because it spans the full color range
observed in these clusters.  The bluest stars in the NGC~6388 BH
population appear to be analogs of the BH stars in the metal-poor
clusters, while the reddest stars in the NGC~6388 BH population appear to be
analogs of the BH stars in the NGC~6441 population.  Moreover, the entire BH
population in NGC~6388 appears to be contiguous.

In some massive clusters, all of the EHB stars may have originated in
a sub-population with enhanced helium.  For example, in NGC~2808,
nearly all of the EHB stars falling within the $Y=0.23$ canonical
locus (Figure 6; {\it pink shaded region}) also fall within the
$Y=0.4$ canonical locus (Figure 6; {\it green shaded region}); such
stars are normal in the sense that they are not flash-mixed.  However,
there is a dearth of stars falling in the brighter part of the
$Y=0.23$ EHB locus, which may indicate that the dominant $Y=0.23$
population in this cluster only produced BHB and red HB stars.  In
NGC~2419, the situation is even more extreme.  Few EHB stars fall in
either the $Y=0.23$ or $Y=0.4$ canonical loci, and thus all of the EHB
stars in this cluster may have been produced via flash-mixing (see
Figure 7).  As with NGC~2808, the dominant $Y=0.23$ population in
NGC~2419 may only have produced BHB and red HB stars.

\subsection{The Case for Flash Mixing}

The faint luminosities observed in the BH stars can be reproduced by
models that undergo flash mixing during a late helium-core
flash on the WD cooling curve.  Helium enhancements may also play an
important role in the formation of BH stars, primarily
because a higher helium abundance lowers the MSTO mass at a given
age so that less mass loss is required on the RGB to
populate the hot HB.  Helium enhancements also depress the
luminosity somewhat in canonical EHB models,
but not by the $\sim$1~mag of the faintest BH stars.

Several recent studies have argued against flash mixing as the origin
of the BH stars.  For example, Lee et al.\ (2005) attribute the BH
stars to the evolution of a population with a high initial helium
abundance ($Y=0.4$).  However, as we demonstrate in \S4.2, an
enhancement in $Y$ cannot by itself explain the BH phenomenon.  While an
enhanced $Y$ decreases the EHB luminosity and increases the BHB
luminosity (thus providing a hook at the hot end of the HB),
one cannot arbitrarily adjust the cluster reddening and distance
to match the theoretical HB locus
to an observed CMD without regard to the relation between the BHB,
the red HB, and the MSTO.  Furthermore, the models of Lee et
al.\ (2005) do not predict any carbon enhancement in the BH stars, while
optical spectra of BH stars in $\omega$ Cen and NGC~2808 do suggest
such an enhancement may be present (Moehler et al.\ 2004, 2007) -- an
enhancement naturally produced by flash-mixing.

D'Antona et al.\ (2010) go further, postulating that MS
stars born at $Y\approx 0.4$ could undergo deep mixing on the RGB that
elevates their helium abundance to $Y\approx 0.8$ before arrival on the
EHB.  The mechanism that drives this deep mixing is unclear, unlike
the mechanism for flash mixing that follows from the basic
stellar structure equations and which is the inevitable outcome
of a helium-core flash on the WD cooling
curve.  However, the effect of the D'Antona et al.\ (2010)
``RGB mixing'' is to generate EHB stars with very high helium abundances
that are hotter than the canonical HB, and in this way their models
are similar to those produced by flash mixing.  An important distinction,
however, is that their RGB mixing precedes the helium-core flash, and thus
does not produce an enhancement in the carbon abundance.  

Castellani et al.\ (2006) have argued that canonical HB models can
reach the high $T_{\rm eff}$ observed for BH stars, but their EHB
stars reach a maximum $T_{\rm eff}$ of $\sim$32,000~K.  This is much
less than that found by Moehler et al.\ (2004, 2007) for the BH stars
($>$35,000~K), and much less than the temperatures reached by either flash
mixing or RGB mixing models.  Furthermore, canonical HB models would
not produce any enhancements in helium or carbon, as observed for such stars
(Moehler et al.\ 2004, 2007).  

In Paper II, we found that a delayed
helium-core flash could sometimes lead to a more shallow form of flash
mixing, where the envelope hydrogen is only mixed into the outer
layers of the core.  When this happens, flash-mixed stars can retain
some residual envelope hydrogen, and consequently would be distributed
closer to the canonical EHB.  
Our models imply that shallow mixing may be a relatively rare
occurrence, especially at low metallicities.  Our CMDs do not show any
systematic variation in the number of BH stars falling close to
the canonical locus, suggesting that shallow mixing is not important
in the metallicity range covered by our clusters.  It may be, however,
that we simply do not have enough BH stars to detect such a trend.
With the current sample, it appears that the BH stars in each cluster
fall up to $\sim$1~mag below the canonical HB, regardless of metallicity.

The sdB and sdO stars of the Galactic field are the analogs of the EHB
and post-EHB stars found in clusters.  One would expect some of these
stars to also show evidence of flash mixing.  This expectation
motivated far-UV spectroscopy of helium-rich sdB stars in the Galactic
field using the {\it Far Ultraviolet Spectroscopic Explorer}, to see
if these stars exhibit the helium and carbon abundance enhancements 
expected from
flash mixing.  Those observations do in fact lend support to the flash
mixing scenario.  The spectrum of PG1544+488 (archetype of the He-sdB
class) shows incredibly strong carbon lines, indicating a surface that is
$\approx 2$\% carbon by mass (Paper II).  

Further insight into the nature of the BH stars
may come from upcoming far-UV spectroscopy of NGC~2808.
In the current observing cycle, we will use the Space Telescope Imaging
Spectrograph on {\it HST} to obtain spectra of BH stars, normal EHB stars,
and BHB stars selected from the NGC~2808 CMD (Figure 5).  With these
spectra, we will be able to search for the atmospheric signatures of flash mixing
(e.g., enhanced carbon, diminished H) and perhaps shed light on the origin
of the wide color range exhibited in BH stars.

Our understanding of the BH phenomenon is far from complete.  None of
the models discussed in this paper (canonical at standard $Y$,
canonical at enhanced $Y$, or flash-mixed) are able to explain 
the red colors found in some or all of the subluminous stars in
each cluster.  This is particularly true in the two metal-rich clusters in
our sample.  Furthermore, the near-total lack of normal EHB stars in
half of our clusters is puzzling.  If we consider the categories of HB
stars (red HB, BHB, normal EHB, BH) as a sequence governed by cluster
parameters (mass loss, helium abundance, cluster mass, etc.), why do
some clusters (NGC~2419, NGC~6388, and NGC~6441) exhibit BHB stars and
BH stars but few normal EHB stars?  The dearth of such stars may imply
a gap in one or more cluster parameters, but the mechanism for such a
gap is unclear.

\subsection{Why Massive Clusters?}
 
A remarkable feature of all clusters either known or suspected to host
BH stars is that they are among the most massive clusters in the
Galaxy.  Dieball et al.\ (2009) also noted that BH stars only appear
to exist in massive clusters, and suggested that this could simply be
a statistical effect, because ``intrinsically rare objects can only be
found by searching a sufficiently large number of stars.''  To examine
this possibility, Dieball et al.\ (2009) apply a statistical test by
comparing the locations of clusters with and without significant BH
populations within the cumulative mass distribution of the clusters in
their sample.  Although Dieball et al.\ (2009) state that their
analysis is ``effectively combining many low-mass clusters to form
aggregate high-mass clusters,'' their test, in fact, ignores the
actual numbers of BH stars in each cluster, and instead only considers
whether clusters do or do not host significant numbers of BH stars.
This is a difficult test, given the small number of clusters in their
comparison.  Although Dieball et al.\ (2009) suggested that it would
be worthwhile to investigate how the numbers of BH stars scale with
cluster properties, they did not actually perform this test due to the
non-uniformities in the extant data.  A uniform data sample would be
important if one were trying to address the total size of the BH
population as a function of cluster mass.  However, the Dieball et
al.\ (2009) sample already contains sufficient data to explore the
relationship between BH population size and cluster mass.  If the UV
imaging is sufficient to produce a statistically significant sample of
EHB stars in each cluster, one can ask if a significant fraction of
the EHB stars are BH stars, and can then compare this fraction in
clusters of different masses.  This comparison is meaningful even if
different clusters are observed in distinct ways (e.g., differences in
region sampled, bandpasses used, depth); the important point is to
observe a large enough sample of EHB stars to determine the relative
numbers of normal and subluminous (BH) EHB stars.

We refer to Table 2 of Dieball et al.\ (2009).  Their table lists 5
clusters hosting significant numbers of BH stars (NGC~2419, NGC~2808,
$\omega$ Cen, NGC~6388, and NGC~6715).  All of these clusters have
masses ranging from 1.42$\times$10$^6$~$M_\odot$ (NGC~2808) to
3.35$\times$10$^6$~$M_\odot$ ($\omega$ Cen).  We would add NGC~6273 to
this list; because it is 0.2~mag fainter in $M_V$ than NGC~2808
(Harris 1996), its mass is $\approx$1.2$\times$10$^6$~$M_\odot$.
Table 2 of Dieball et al.\ (2009) also lists 11 clusters that do not
have significant numbers of BH stars.  Of course, besides these 11
clusters, there are many more low-mass and intermediate-mass clusters
without known BH stars, but these 11 were considered by Dieball et
al.\ (2009) to have sufficient imagery to rule one way or the other.
Among that list of 11 clusters is NGC~6441, with a mass of
1.57$\times$10$^6$~$M_\odot$; we would move it to the list of
clusters hosting BH stars.  Also in that list of 11 clusters is
NGC~104 (47 Tuc); it is massive (1.5$\times$10$^6$~$M_\odot$), but it
is the only cluster in the entire table that has no BHB stars (its HB
population is entirely red).  The total mass of the 9 remaining clusters in 
this list is 5.1$\times$10$^6$~$M_\odot$.  The stellar
mass in this ensemble of 9 clusters exceeds the mass in any individual
cluster, and is twice the mass of any cluster we have explored in our
own study.  Although this ensemble is comprised of 9 clusters hosting
BHB stars and in some cases (e.g., M13, M80, NGC~6752) substantial 
populations of EHB stars, this ensemble has few if any BH stars (possibly 
3--5; Dieball et al.\ 2009; Sandquist et al.\ 2010).  
If finding BH stars simply required the searching 
of enough stellar mass, we would expect to find tens of BH stars in the
ensemble mass of 9 clusters totaling 5.1$\times$10$^6$~$M_\odot$.  The
implication is that any cluster hosting BHB stars will also host BH
stars if its mass exceeds $\sim$1.2$\times$10$^6$~$M_\odot$.
However, the fraction of EHB stars that are subluminous varies from cluster 
to cluster in our sample, and that fraction does not appear to correlate
with either cluster mass or metallicity, implying additional factors 
are at work (including perhaps the dynamical history of interactions
between the cluster and the Galaxy).  In half of our sample
(NGC~2419, NGC~6388, and possibly NGC~6441), virtually all of the EHB 
stars appear to be BH stars, with very few stars on the
canonical EHB, whereas in the remaining three clusters
(NGC~6273, NGC~6715, and NGC~2808), the split between
normal and subluminous EHB stars appears to be roughly equal. 

Note that low-mass clusters can have large samples of EHB stars
without having any BH stars.  Consider NGC~6752, which is one of the 9
clusters discussed above that hosts no BH stars; it is listed in Table
2 of Dieball et al.\ (2009) with a mass of
0.32$\times$10$^6$~$M_\odot$.  For NGC~6752 (Landsman et al.\ 1996)
and NGC~2808 (Paper I), there exist CMDs that include a far-UV
bandpass.  In the case of NGC~6752, the CMD is sampling a less massive
cluster outside of the core, while in the case of NGC~2808, the CMD is
sampling a more massive cluster within the core.  Despite these
distinctions, the samples of EHB stars and their descendents are
comparable in size.  In each CMD, there are $\sim$80 HB stars hotter
than 16000~K, along with four post-EHB stars -- i.e., the
shorter-lived AGBM stars.  However, for the case of NGC~2808, 46 of
the EHB stars are BH stars (lying well below the canonical EHB), while
in NGC~6752, none of them are. Given that the size of the EHB samples
are approximately the same, the number of BH stars in the two clusters
are discrepant at a high statistical significance.  This argument is
not restricted to NGC~6752; other low-mass clusters, such as M13 and
M80 (Ferraro et al.\ 1998) host dozens of EHB stars but apparently few
if any BH stars (M13 may host 2 BH stars; Sandquist et al.\ 2010).
The large numbers of EHB stars in both the low-mass cluster sample and
the high-mass cluster sample provide strong support for the notion
that the fraction of EHB stars that are subluminous (i.e., BH stars)
is nearly zero in low-mass clusters and $\sim$50--100\% in high-mass
clusters.

There are several possible reasons why BH stars may only appear
in massive globular clusters.  One possibility is that only massive
clusters can retain the helium-rich ejecta from their first generation of
stars.  Besides this helium enrichment, such self-pollution might account
for many of the abundance variations in O, Na, Mg and Al observed in
MS and RGB stars (see Gratton et al.\ 2004 and references
therein).  

An alternative hypothesis is that massive clusters have many close
binaries, which produce the high mass loss needed for flash mixing.
Of the massive clusters known to host BH stars, none are
core-collapsed, and all are of intermediate central concentration
(Harris 1996), so that one might expect a significant fraction of
surviving binaries (Ivanova et al.\ 2005).  However, the lack of a
radial gradient in the NGC~2808 BH population argues against a
binary origin (Bedin et al.\ 2000).

\acknowledgements

Support for proposal 10815 is provided by NASA through a grant from
STScI, which is operated by AURA, Inc., under NASA contract NAS
5-26555.  We thank the anonymous referee for comments that helped
improve the clarity of this paper.

\end{document}